\documentclass[prl,aps,twocolumn,showpacs]{revtex4}

\usepackage{dcolumn}
\usepackage{amsmath}
\usepackage{graphics}

\begin{document}

\title{Light scattering and localization in an ultracold and dense atomic
system}

\author{I.M. Sokolov, M.D. Kupriyanova${}^{1}$, D.V. Kupriyanov}%
\affiliation{Department of Theoretical Physics, State
Polytechnic University, 195251, St.-Petersburg, Russia}%
\email{kupr@dk11578.spb.edu}%

\altaffiliation[$^{1}$currently at ]{V.A. Fock Physics Institute,
St.-Petersburg University,
198504 Stary Petershof, St.-Petersburg, Russia}%

\author{M.D. Havey}%
\affiliation{Department of Physics, Old Dominion University,
Norfolk, VA 23529}%

\date{\today }

\begin{abstract}
The quantum optical response of high density ultracold atomic
systems is critical to a wide range of fundamentally and
technically important physical processes. These include quantum
image storage, optically based quantum repeaters and ultracold
molecule formation. We present here a microscopic analysis of the
light scattering on such a system, and we compare it with a
corresponding description based on macroscopic Maxwell theory.
Results are discussed in the context of the spectral resonance
structure, time-dependent response, and the light localization
problem.
\end{abstract}


\maketitle%

%% 34.50.Rk, 34.80.Qb - Laser modified scattering and reactions
%% 42.50.Ct - Quantum description of interaction of light and
%%            matter, related experiments

Early studies of formation and dynamics of ultracold atomic
samples were largely focused on obtaining the high density and low
temperatures necessary to attain Bose-Einstein Condensation in
atomic gas samples \cite{Pethick}. At the same time, a large
number of other research areas emerged from studies of ultracold
atomic gas samples.  Among these, there are several where physical
processes are importantly modified at high atomic density. These
include efforts in ultracold molecule formation \cite{Wester},
image storage in high-optical depth samples \cite{Howell1}, and
light storage and manipulation for possible atomic-physics based
quantum repeaters \cite{Bouwmeester} and other quantum information
applications. Another little explored area which may have
significant impact in a range of scientific or technical areas is
study of quantum optical processes at high density $n_0$, where
$n_0\lambdabar^3$ $\sim$ 1 \cite{Dicke}. Under these conditions,
disorder-dependent phenomena such as an Anderson-type
\cite{Anderson} localization, random lasing \cite{HuiCao1}, and
other mesoscopic phenomena \cite{AkkermanBook,Sheng,Gero} may
appear.  In these cases, the quantum optical response of the
medium is of critical importance to dressing and probing the
physical processes involved.

In this letter we are concerned with the collective dynamics of
light and a sample of ultracold and high-density atoms. Under the
conditions that $n_0\lambdabar^3\gtrsim$ 1 the interaction of
atomic dipoles via longitudinal and transverse electric fields
strongly interferes with the scattering process. The atomic
dipoles cannot be considered as independent secondary sources of
the scattered waves freely propagating through the sample, as
usually takes place for an optically dilute system. Here we
present theoretical discussion of the scattering process based on
two different and complementary approaches. In the first, we use a
self-consistent description of the atomic sample in the spirit of
the Debye-Mie model for a macroscopic spherical scatterer
consisting of a homogeneous and dense configuration of atomic
dipoles. We apply a self-consistent calculation of the macroscopic
permittivity based on its relevant statistical expression given by
the Kubo formula \cite{LaLfIX}. In this model the scattered light
is restored as secondary waves created by randomly distributed
atomic dipoles, whose reradiation should be properly expanded in
the spherical modes of the Debye-Mie problem. In the second
approach we make exact numerical analysis of the completely
quantum-posed description of the single photon scattering problem.
The atoms, the incoming photon and the field modes participating
in the process are considered in a second quantized formalism.
This analysis gives us the exact value of the scattering amplitude
and in particular shows the correct description of the cross
section in those spectral domains where the Debye-Mie model fails
or is not irreproachably applicable. We show the presence of
collective micro-cavity structure, created by the atomic system,
where a number of resonant modes have a sub-radiant nature, and
discuss their physical substance in the context of light
scattering and diffusion. The results have important implications
for studies of light localization, and for manipulation of single
and few photon states in ultracold atomic gases.

Light transport through a disordered system of atomic dipoles can
be discussed in the Maxwell theory as a combination of the process
of coherent propagation, described by configuration averaged
Maxwell equations, and incoherent losses initiated by scattering
from mesoscopically scaled permittivity fluctuations. In such a
formalism one can visualize the light state as a local plane wave
existing on a spatial scale $\sim \lambda$ and coherently
interacting with the system of atomic dipoles located in a
relevant mesoscopic volume. The dipoles interacting with the field
are locally indistinguishable and the interaction process is
assumed to be well approximated by the cooperative coherent
response \cite{LaLfIX} for the sample susceptibility
$\chi(\omega)$ at frequency $\omega$. Then the permittivity
$\epsilon(\omega)=1+4\pi\chi(\omega)$ can be extracted via a
self-consistent approximation of the collective dipole dynamics
driven by the transverse electric field and modified by the
Lorentz-Lorenz effect coming from interference of proximal atomic
dipoles. If incoherent losses have only a radiative nature and
atoms equally populate the ground state Zeeman sublevels we obtain
a self-consistent expression for $\epsilon(\omega)$ for an
infinite sample
\begin{equation}
\epsilon(\omega)=\frac{1-\frac{8\pi
n_0}{3\hbar}\frac{|d_{F_0F}|^2}{3(2F_0+1)}%
\frac{1}{\omega-\omega_0+i\sqrt{\epsilon(\omega)}\gamma/2}}%
{1+\frac{4\pi n_0}{3\hbar}\frac{|d_{F_0F}|^2}{3(2F_0+1)}%
\frac{1}{\omega-\omega_0+i\sqrt{\epsilon(\omega)}\gamma/2}}%
\label{1}%
\end{equation}
This equation can be analytically solved and $\epsilon(\omega)$
expressed in terms of well defined external parameters, such as
the natural decay rate $\gamma$ and $n_0$. The reduced matrix
element $d_{F_0F}$ for a given dipole transition can be also
expressed by $\gamma$. We consider here a closed transition, such
that for hyperfine levels only the selected lower and upper states
with angular momenta $F_0$ and $F$ respectively can be
radiatively. The above result can be applied to calculation of the
scattering cross section of light by a spherical sample of large
radius $a$ ($a\gg\lambdabar$) in the standard formalism of the
Debye-Mie problem.
\begin{eqnarray}
Q_S&=&\frac{\pi}{2k^2}\sum\limits_{J=1}^{\infty}(2J+1)%
\left[\left|1-S_J^{(e)}\right|^2+\left|1-S_J^{(m)}\right|^2\right]\phantom{2.1}%
\nonumber\\%
Q_A&=&\frac{\pi}{2k^2}\sum\limits_{J=1}^{\infty}(2J+1)%
\left[1-\left|S_J^{(e)}\right|^2+1-\left|S_J^{(m)}\right|^2\right]\phantom{2.1}%
\label{2}%
\end{eqnarray}
where the scattering matrix components for the $TM$ and $TE$ modes
are respectively given by
\begin{eqnarray}
S_J^{(e)}\!\!&=&\!-\frac{\epsilon(\omega)j_J(ka)[rh_J^{(2)}(\frac{\omega}{c}r)]'_{r=a}%
\!-\!h_J^{(2)}(\frac{\omega}{c}a)[rj_J(kr)]'_{r=a}}%
{\epsilon(\omega)j_J(ka)[rh_J^{(1)}(\frac{\omega}{c}r))]'_{r=a}%
\!-\!h_J^{(1)}(\frac{\omega}{c}a)[rj_J(kr)]'_{r=a}}%
\nonumber\\%
S_J^{(m)}\!\!&=&\!-\frac{j_J(ka)[rh_J^{(2)}(\frac{\omega}{c}r)]'_{r=a}%
\!-\!h_J^{(2)}(\frac{\omega}{c}a)[rj_J(kr)]'_{r=a}}%
{j_J(ka)[rh_J^{(1)}(\frac{\omega}{c}r))]'_{r=a}%
\!-\!h_J^{(1)}(\frac{\omega}{c}a)[rj_J(kr)]'_{r=a}}%
\nonumber\\%
&&\label{3}%
\end{eqnarray}
Here $j_J(\ldots)$, $h_J^{(1)}(\ldots)$ and $h_J^{(2)}(\ldots)$
are spherical Bessel functions of $J$-th order, and
$k=\sqrt{\epsilon(\omega)}\,\omega/c$. $Q_S$ is the elastic part
of the cross section responsible for the coherent scattering of
light from the sample boundary. The absorption part $Q_A$ is
responsible for diffusely scattered light via the incoherent
channel and the sum $Q_0=Q_S+Q_A$ is the total cross section for
the entire scattering process. Light emerging the sample via
incoherent channels can be recovered in the Maxwell theory by
considering secondary and multiply scattered waves generated by
fluctuations of $\chi(\omega)$. This can be simulated by a
Monte-Carlo scheme and crucially requires that the dipole sources
generating these waves be indistinguishable on a mesoscopic
distance $\sim\lambda$. Then secondary and multiply scattered
waves can be simulated through the time dependence of the sample
fluorescence when the excitation light is turned off. The analysis
of the transient process and time dependence of the fluorescence
is simplified by a diffuse approximation if the extinction length
for the field penetration inside the sample is $\gg\lambda$.

We now turn to the quantum-posed description of the photon
scattering problem.  There photon scattering on an atomic system
is expressed by the following relation between the total cross
section and the $T$-matrix
\begin{equation}
Q_0=\frac{{\cal V}^2}{\hbar^2
c^4}\frac{\omega'^2}{(2\pi)^2}\int\sum\limits_{\mathbf{e}'}%
\left|T_{g\mathbf{e}'\mathbf{k}',g\mathbf{ek}}(E_\mathrm{i}+i0)\right|^2d\Omega%
\label{4}%
\end{equation}
where ${\cal V}$ is a quantization volume. Here we keet only the
Rayleigh channel and assumed that the atomic system is described
by the same ground state $g$ before and after the scattering
process, which includes the averaging over initial and sum over
final Zeeman states. The initial energy of the entire system
$E_\mathrm{i}$ is given by $E_\mathrm{i}=E_g+\hbar\omega$ and the
incoming and outgoing photons have the same frequency
$\omega'=\omega$. We consider below the simplest relevant example
of a "two-level" atom, which has only one sublevel in its ground
state and three Zeeman sublevels in its excited state, such that
$F_0=0$ and $F=1$ and in an isotropic situation the total cross
section does not depend on the momentum direction and polarization
of the incoming photon.

The $T$-matrix is expressed by the total Hamiltonian $H=H_0+V$ and
by its interaction part $V$ as
\begin{equation}
T(E)=V+V\frac{1}{E-H}V.%
\label{5}%
\end{equation}
In the rotating wave approximation the internal resolvent operator
contributes to (\ref{4}) only by being projected on the states
consisting of single atom excitation, distributed over the
ensemble, and the vacuum state for all the field modes. Defining
such a projector as $P$ the projected resolvent
\begin{equation}
\tilde{R}(E)=P\frac{1}{E-H}P%
\label{6}%
\end{equation}
performs a $3N\times 3N$ matrix, where $N$ is the number of atoms.
For the dipole-type interaction between atoms and field this
projected resolvent can be found as the reversed matrix of the
following operator
\begin{equation}
\tilde{R}^{-1}(E)=P\left(E-H_0-VQ\frac{1}{E-H_0}QV\right)P%
\label{7}%
\end{equation}
where the complementary projector $Q=1-P$, operating in the
self-energy term, can generate only two types of intermediate
states: a single photon $+$ all the atoms in the ground state; and
a single photon $+$ two different atoms in the excited state and
others are in the ground. For such particular projections there is
the following important constraint on the dipole-type interaction
$V$: $PVP=QVQ=0$. Due to this constraint the series for the
reversed resolvent (\ref{7}) is expressed by a finite number of
terms, which would be not the case in a general situation
\cite{ChTnDRGr}. The resolvent $\tilde{R}(E)$ can be numerically
calculated and, for an atomic system consisting of a macroscopic
number of atoms, when $N\gg 1$, the microscopically calculated
cross section (\ref{4}) can be compared with the Debye-Mie theory
results (\ref{2}). Both the microscopic and self-consistent
approaches have no fitting parameters. Compared with the
mesoscopic approach, the essential difference is that in the
microscopic approach {\it any type of collective atomic
excitations} can contribute, which are {\it not responsibly
cooperative}. One can then expect significant difference in their
predictions for the cross section and for the fluorescence
behavior. The time dependence of the fluorescence signal can be
built up via Fourier expansion of the outgoing photon wavepacket
with the $S$-matrix formalism.

\begin{figure}[t]
\includegraphics{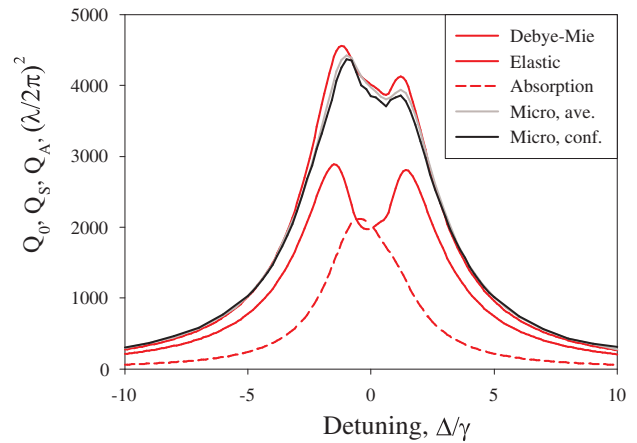}%
\caption{The spectral dependence of the total $Q_0$, elastic
$Q_S$, and absorption $Q_A$ cross sections for an atomic sample of
size $a=25\lambdabar$ and density $n_0\lambdabar^3=0.02$. The
configuration averaged microscopic result, shown as a gray solid
line, is evidently the same as the data taken for particular
atomic configuration, shown as a black line. The results of a
mesoscopic self-consistent approach (red curves) reproduce the
exact microscopic spectra. Color online.}
\label{fig1}%
\end{figure}

\begin{figure}[t]
\includegraphics{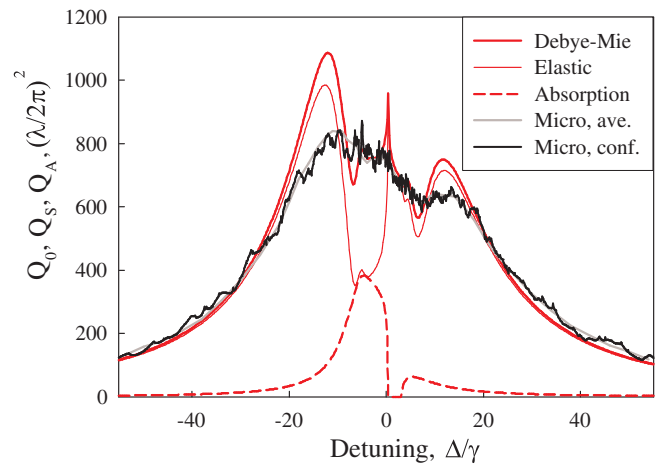}%
\caption{Same as Fig.\ref{fig1}, but for a sample size
$a=10\lambdabar$ and density $n_0\lambdabar^3=0.5$. The results of
the self-consistent mesoscopic approach (red curves) perceptibly
differ from the exact microscopic spectra. In turn, the
microscopic calculation given for a particular configuration
(black curve) indicates speckle micro-cavity structure generated
by the resolvent poles. Color online.}
\label{fig2}%
\end{figure}

Below we consider the spectral dependence of the cross sections,
calculated in both the self-consistent mesoscopic model and in an
exact microscopic approach. Our central idea is to follow how this
dependence is modified when the dimensionless density of atoms
$n_0\lambdabar^3$ is varied from smaller to greater values. In
Fig. \ref{fig1} we reproduce such a spectral dependence when the
density of atoms is small, $n_0\lambdabar^3=0.02$. There is
excellent agreement between the Debye-Mie and microscopically
calculated data and the microscopic result is insensitive to
configuration averaging over random atomic distributions. This
justifies that only cooperative modes, which allow the macroscopic
Maxwell description, contribute to the scattering process. As a
consequence the long term time dynamics of the fluorescence signal
can be very well approximated by a Holstein mode. Such a spectral
and temporal behavior is typical for dilute atomic systems as was
verified by our numerical simulations done for the atomic
ensembles of different sizes and consisting of different numbers
of atoms. While varying the density to greater values
$n_0\lambdabar^3\sim n_c\sim 0.09$ the solution of the
self-consistent equation (\ref{1}) turns the permittivity to
negative values in a part of the spectrum where $\epsilon'<0$ and
$\epsilon''=0$. As is well known in, e.g. plasma physics
\cite{LaLfIX} the negative permittivity can be associated with a
forbidden spectral zone, where the radiation cannot penetrate
inside the system and can exist only in the form of a surface
light-matter wave. In the limit of even higher density samples,
the light undergoes mainly surface scattering and the absorption
(incoherent/diffusion scattering) channels suppressed. This is
illustrated in Fig. \ref{fig2} by those spectral dependencies of
the cross section, which were calculated in the Debye-Mie model
for $n_0\lambdabar^3=0.5$. In contrast with similar results in
Fig. \ref{fig1} the contribution of the elastic channel now
dominates such that a smaller portion of light can penetrate the
sample. As a consequence, it is expected that fewer atomic
excitations should contribute to the time dependent fluorescence
decay and the self-consistent model predicts an even faster decay
than would take place in a similar dilute system.

\begin{figure}[t]
\includegraphics{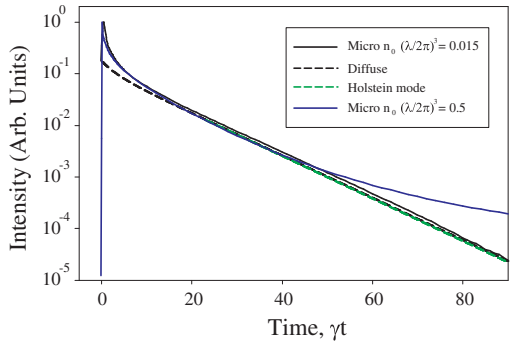}%
\caption{Time dependent fluorescence decay for samples of
different densities and with relatively equal optical depths. The
dilute system has a sample radius $a=27\lambdabar$ and density
$n_0\lambdabar^3=0.015$, and for the dense system these
characteristics are given by $a=5\lambdabar$ and
$n_0\lambdabar^3=0.5$. Color online.}
\label{fig3}%
\end{figure}

The situation can be differently posed if it is considered from a
microscopic point of view. Then such a dense atomic vapor can be
relevantly described as a micro-cavity system built up in an
environment of randomly distributed atomic scatterers. Any
distribution creates a specific quantization problem for the
incoming field, whose mode structure can be properly defined in
terms of standard scattering theory. The main difficulty for the
quantization procedure is description of the complicated structure
of the resonance states. The resonances are described by the
resolvent poles and can be specified by various superpositions of
atomic states, which transform the reversed projected resolvent
(\ref{7}) to diagonal form. The micro-cavity structure manifests
itself by finely resolved speckle dependence in the scattering
spectrum and the scattering process becomes extremely sensitive to
the spatial configuration of atomic scatterers, as illustrated in
Fig. \ref{fig2}. Some of these resonance states have a subradiant
nature and manifest themselves via significantly slowed long time
decay of the fluorescence in comparison with a classical Holstein
mode. Fig. \ref{fig3} illustrates the difference in the
fluorescence decay for dilute and dense atomic systems initially
excited by a short probe pulse. The external parameters for the
systems are such that the optical depth for both the systems is
nearly the same. For a dilute system the long term asymptote,
extracted through exact microscopic calculations, is well
described by a diffuse Holstein-type mode evaluated via the sample
macroscopic characteristics. For the dense system there is evident
deviation of the asymptotic behavior from such behavior. Such a
deviation should be associated with the presence of sub-radiant
resonance states in the resolvent poles.

The analogy of sub-radiant states with a localization process,
usually discussed as a multiple wave scattering problem, suggests
evolution with density of the qualitative scattering properties of
the atomic gas. The discussed characteristics, such as the
spectral dependence of the sample cross-section or the
time-dependent fluorescence, have no direct relations with static
thermodynamic properties of the system, which could be evaluated
based on its partition function. However, they describe how the
system in equilibrium responds to optical excitation near the
atomic resonance transition. The atomic transition is dressed by
cooperative dipole-field interactions and at higher densities, the
disordered atomic system reveals a micro-cavity structure. Such a
micro-cavity has unique properties for any particular
configuration and its mode splitting is competitive with the mode
decay rate. The modification of the dynamic response to an
external field with density is an intrinsic property of the
system, which may be visualized in terms of a cross over or a
change of phase. In this sense our calculations show that the
transformation of the system behavior from one of individual,
independent atomic scatterers to the cooperatively organized
micro-cavity structure mainly develops within a narrow and
critical density zone. The configuration sensitive speckle
structure of the spectral cross section manifests itself in those
conditions when the self-consistent permittivity becomes negative
in part of the spectrum. The description of the atomic system with
the macroscopic Maxwell theory preferably yields surface
scattering of light in this case. More precise microscopic
description shows that part of the excitation can penetrate the
system and be converted into a long-time decay of the fluorescence
signal.

We appreciate the financial support by NSF (Grant No.
NSF-PHY-0654226) and INTAS (Project No 7904).

\end{document}